\documentclass[useAMS,usenatbib]{mn2e}

\title[Angular momentum of collided preplanetesimals]
  {The angular momentum of colliding rarefied
preplanetesimals and the formation of binaries }
\author[S. I. Ipatov]
  {S. I. ~Ipatov$^{1,2}$\thanks{E-mail: siipatov@hotmail.com}
   \\
  $^1$Catholic University of America, Department of Physics, Washington DC, 20064, USA\\
  $^2$Space Research Institute, Moscow, Russia}
\date{Submitted}

\pagerange{\pageref{firstpage}--\pageref{lastpage}} \pubyear{2009}

\def\LaTeX{L\kern-.36em\raise.3ex\hbox{a}\kern-.15em
    T\kern-.1667em\lower.7ex\hbox{E}\kern-.125emX}

\begin{document}

\label{firstpage}

\maketitle

\begin{abstract}
This paper studies the mean angular momentum associated with the collision of 
two celestial objects in the earliest stages of planet formation. Of primary concern is the scenario of two rarefied preplanetesimals (RPPs) in circular heliocentric orbits. The theoretical results are used to develop models of binary or multiple system formation 
from RPPs, and explain the observation that a greater fraction of 
binaries originated 
farther from the Sun. At the stage of RPPs, small-body satellites
can form in two ways: a merger between RPPs can have two centers of contraction, or the formation of satellites from a disc around the primary or the secondary.
Formation of the disc can be caused by that the angular momentum 
of the RPP formed by the merger is greater than the 
critical angular momentum for a solid body. One or several
satellites of the primary (moving mainly in low-eccentricity orbits) can be formed from 
this disc at any separation less than the Hill radius.
The first scenario can explain a system such as 2001 QW$_{322}$ where the two components have similar masses but are separated by a great distance. In general, any values 
for the eccentricity 
and inclination of the mutual orbit are possible. Among discovered 
binaries, the observed angular momenta are smaller than the typical angular momenta expected for identical 
RPPs having the same total mass as the discovered binary and encountering each other in circular heliocentric orbits. 
This suggests that the population of RPPs underwent some contraction before mergers became common.
\end{abstract}


\section{Introduction}
\subsection{The case for rarefied preplanetesimals}
From the 1950's to the 1980's, many authors (e.g. Safronov 1969; Goldreich \& Ward 1973)
believed that planetesimals
originated from {\em rarefied dust condensations} (RDCs): overdense regions of the disc that are gravitationally bound, and may contain solid particles or bodies, but are still gaseous or diffuse. During this period, the mechanisms behind RDC formation and the coagulation of larger 'secondary condensations' were studied by several authors (e.g. Safronov 1969, 1995; 
Eneev \& Kozlov 1981, 1982; Pechernikova \& Vityazev 1988). The duration of the condensation stage was thought to be longer at greater distances from the Sun, where the RDCs have lower densities. According to an early model by Safronov (1967, 1969), the time required for RDCs to form into solid bodies was about 10$^4$ yr in the terrestrial zone and 10$^6$ yr at the distance of Jupiter. The corresponding increases in mass were 10$^2$ and 10$^3$ from the initial values. Safronov concluded that while the rotation of initial RDCs impedes contraction, they also become denser when they combine.

In the models of RDC contraction considered by Myasnikov \& Titarenko 
(1989, 1990), collisions between
condensations were not taken into account. 
The times needed for RDCs to become solid planetesimals
could exceed several million years, depending on 
the optical properties of the dust and gas and the
type and concentration of short-lived radioactive isotopes in RDCs. Depending on the concentration, the times varied by a factor of more than 10 and could exceed 10 Myr
at the concentration greater than 0.02.  

In the 1990's, scenarios of planetesimal formation through the
gravitational instability of RDCs were frustrated by the phenomenon of self-generated mid-plane turbulence. Models involving the hierarchical
accretion of planetesimals from smaller solid objects became more popular (e.g. Weidenschilling 2003). In recent
years, however, new arguments 
(e.g. Makalkin \& Ziglina 2004; 
Johansen et al. 2007; Cuzzi, Hogan \& Shariff 2008; Lyra et al. 2008;
Johansen,  Youdin \& Mac Low 2009)
have arisen in favour of RDCs, or as they can be also called
{\em rarefied preplanetesimals} (RPPs)
(condensations, clusters, and clumps). Unlike earlier models, however, the overdense
regions could include metre-sized boulders and smaller solid bodies. Some results from these more recent models are summarised below.

Makalkin \& Ziglina (2004) showed that after the
subdisc reaches a certain critical density, its inner, equatorial layer could 
be gravitationally unstable. This is possible because the inner disc, unlike 
the two subsurface layers, contains no shear turbulence.  In their model, 
RDCs form trans-Neptunian objects (TNOs) with diameters up
to 1000 km in $\sim$$10^6$ yr. 

Johansen et al. (2007) found that gravitationally bound
clusters can form with masses comparable to dwarf planets.
They determined that peak densities in the clusters approached
$10^4\rho_{gas}$ (where $\rho_{gas}$ is the average gas density) after
only seven orbits. 
Lyra et al. (2008) noted that strong drag forces in the disc might delay
gravitational collapse of the clusters. 
Cuzzi et al.
(2008) showed how self-gravity could stabilise dense clumps
of millimetre-sized particles. Such clumps form naturally in 3D
turbulence, forming diffuse but cohesive 'sandpiles' on the order
of 10--100 km in diameter. For chondrules of radius
$r_c=300 \mu$m, they obtained a `sedimentation' timescale of
about 30--300 orbital periods at 2.5 au. (The timescale is proportional to $r_c^{-1}$.) 
Youdin (2008) summarised various mechanisms for
particle concentration in gas discs, including turbulent pressure
maxima, drag instabilities, and long-lived anticylonic vortices.

Ida, Guillot \& Morbidelli (2008) studied the accretion and
destruction of solid planetesimals in turbulent discs, and found
that accretion proceeds only for planetesimals with diameters
above 300 km at 1 au and above 1000 km at 5 au. 
Their analytical arguments were based on fluid dynamical simulations and orbital integrations.
They
concluded that some mechanism must be capable of producing 
Ceres-mass planetesimals on very short timescales. Based on
the observed size-frequency distribution and collisional evolution
of asteroids, Morbidelli et al. (2009) concluded that
the initial planetesimals had to range from one hundred to
several hundreds of km in size, probably up to 1,000 km.

Prior to these new arguments in favour of RPPs, 
I (Ipatov 2001, 2004) had suggested that TNOs
with mildly eccentric orbits ($e<0.3$) and diameters greater 
than 100 km (such as Pluto and Charon) could result from the compression
of large RPPs with semimajor axes $a > 30$ au rather than the accretion of small, solid
planetesimals. I also proposed that some planetesimals with
diameters $d\sim$ 100--1000 km in the feeding zone of the giant planets,
some planetesimals with $d\sim$ 100 km in the terrestrial 
zone, and some large main-belt asteroids could have
formed by direct compression. Some smaller objects (TNOs,
planetesimals, and asteroids) with $d<100$ km could be debris
from large objects, while others could have formed
by direct compression of preplanetesimals.

\subsection{Formation of binaries}

The frequency of binary systems and the ratio of secondary mass 
to primary mass are greater among classical TNOs with inclinations 
$i<5^{\circ}$  than among main-belt asteroids
(Richardson \& Walsh 2006; Knoll 2006; 
www.johnstonsarchive.net/astro/asteroidmoons.html). 

Planetesimal formation models based on the accretion of solid bodies
have put forth several hypotheses on the formation of binaries. 
For example, several papers 
(e.g. Doressoundiram et al. 
1997; Durda et al. 2004; Canup 2005; Walsh 2009)
have been devoted to the mechanism
of catastrophic collisions. Weidenschilling
(2002) studied the collision of two planetesimals
within the sphere of influence of a third body. Goldreich,
Lithwick \& Sari (2002) proposed that the primary body could 
capture a secondary component passing inside its Hill sphere 
due to dynamical friction from surrounding small bodies or 
gravitational scattering by a third large body. Funato et al.
 (2004) studied a model in which a low-mass secondary
component is ejected and replaced by a third
body in a wide but eccentric orbit. Astakhov, Lee \& Farrelly
(2005) studied binary system formation in four-body simulations
with solar tidal effects. Gorkavyi (2008) proposed a
multi-impact model. Six mechanisms of binary formation 
were discussed by Doressoundiram (1997). More
references can be found in papers by Richardson \& Walsh (2006), 
Noll (2006), Petit et al. (2008), Noll et al. (2008), and Scheeres (2009). 

We agree with \' Cuk (2007), Pravec, Harris \& Warner (2007), 
Walsh, Richardson \& Michel  (2008), and 
several
other authors that binary systems with small primaries 
(such as near-Earth objects) arise mainly from the 
rotational breakup of `rubble piles', for example due
to increasing spin from the 
Yarkovsky-O'Keefe-Radzievskii-Paddack (YORP) effect. 
However, I believe that collisions 
could also push small bodies beyond their critical spin limits.

Some scientists believe that the formation of small-body
binaries could be similar to the formation of binary
stars within fragmented discs (Alexander, Armitage 
\& Cuadra 2008). Nesvorny (2008) noted that an excess of
angular momentum prevents the agglomeration of all available
mass into solitary objects during gravitational collapse.

\subsection{Problems studied in this paper}

In contrast with models of binary formation that consider
solid bodies, in 2004 I proposed that a considerable
fraction of trans-Neptunian binaries (including Pluto--Charon) 
could form while rarefied preplanetesimals moving in almost 
circular orbits undergo compression (Ipatov 2004). In 2009, I set forth two 
detailed models of binary formation during this stage
(Ipatov 2009). They are repeated here 
and further discussed in Sections 2.2 and 2.3. These models can 
explain trans-Neptunian binaries, and some asteroid binaries with large primaries
(with diameters of at least 50 km). Small asteroid binaries
could mainly 
arise from collisions between larger solid bodies.

The problem of binary formation from RPPs
has several aspects (e.g. simulations of contraction,
collisions between preplanetesimals). In the present paper, I
pay particular attention to calculating the angular
momentum of two colliding RPPs about
their centre of mass (Section 3). I also compare the predicted
angular momenta to the observed momenta of discovered binary systems 
(Section 4). Other aspects of binary formation
at the preplanetesimal stage (e.g.  mergers
of colliding preplanetesimals) are discussed briefly in Section
5, and may be the subject of future publications.

We show that if RPPs existed during
the period permitting encounters to occur between
preplanetesimals within their Hill spheres, then it is possible
to explain the differences between TNO and asteroid
binaries as well as some of the peculiarities of observed binaries. The 
material presented in Section 3 also helps us better understand the 
angular momenta of planetesimals, trans-Neptunian
objects, and asteroids, although these momenta could change
considerably through gravitational interactions and collisions 
after all bodies have solidified.

\section{SCENARIOS OF BINARY FORMATION DURING THE RAREFIED
PREPLANETESIMAL STAGE}

\subsection{Application of previous solid-body scenarios to preplanetesimals}

At present, most small bodies move 
in orbits with eccentricities exceeding 0.05. Solid planetesimals
would have begun with much smaller eccentricities, but mutual gravitation
tends to increase the eccentricities and inclinations of their orbits (e.g. Ipatov 1988, 2007). 
RPPs, on the other hand, would remain in almost circular orbits
with very small inclinations 
before the moment of formation of a binary. 
The smaller growth of RPP orbital eccentricities 
could be because the stage of preplanetesimals was short,
and circulation of an orbit of an object (e.g. a preplanetesimal
or a planetesimal) of a fixed mass due to the influence of
gas and dust could be greater for a larger size object. Due to
their larger sizes, preplanetesimals collided more often than
solid planetesimals per unit of time.

If we consider almost circular heliocentric orbits, 
the typical minimum distance between the centres of
mass of the objects could be smaller than that between two 
objects entering the Hill sphere from eccentric heliocentric
orbits. Heliocentric orbits with high eccentricities more often 
give rise to hyperbolic relative motion. In contrast,
for nearly circular orbits, the trajectory of relative
motion inside the sphere can be complicated (e.g. Ipatov 1987; 
Greenzweig \& Lissauer 1990; Iwasaki \& Ohtsuki 2007).
  In this case, the objects
can move inside the sphere for a long time, and the relative
distance can change considerably. The role of tidal forces could also be
greater for rarefied objects.

The above discussion supports my belief that models
of binary formation due to gravitational interaction or
collisions occurring inside the Hill sphere of the future primary, 
which have been considered by several authors for solid objects, 
could be even more effective for rarefied preplanetesimals.

\subsection{Binary formation from two centres of contraction}

Rarified preplanetesimals contract over time, and may exhibit centrally concentrated
radial density functions. It is reasonable to suppose that in some cases, the collision of
two such objects results in a rotating system with two centres of contraction.
The end result of gravitational collapse would be a
binary system with almost the same total mass and components
separated by a large distance (such as 2001 QW$_{322}$).
For such a scenario, the values of the eccentricity of the
present mutual orbit of components can have any values less than 1.
 
If the original RPPs were similar in size to their Hill spheres, 
the separation $L$ of the solid binary resulting from the combined system
could range up to the radius $r_H$ of the Hill sphere. More often, however, the
preplanetesimals were much smaller than their Hill spheres.
The separation distance $L$ would then be less than $r_H$. For discovered
binaries with $L/r_p>100$, the ratio $r_s/r_p$ of the radii of
secondary to primary components is greater than 0.5. This
fact may imply that when the masses of two rarified preplanetesimals differ
greatly, only one centre of contraction survives. Note that other authors 
(e.g. Descamps \& Marchis 2008)
have suggested different explanations for the above ratio.

\subsection{Binary formation due to excessive angular momentum}

In my opinion, some binaries could have formed from RPPs 
that obtained more
angular momentum than the maximum possible value for a solid
body. Let $v_s$ denote the velocity of a particle on the surface
of a rotating object, and let $v_{cf}$ be the minimum velocity at
which a particle can leave the surface. As a
rotating rarefied preplanetesimal contracts, some of the material with
$v_s>v_{cf}$ could form a cloud (that transforms into
a disc) that moves around the primary. For a
spherical object, $v_s$ is greatest at the equator and $v_{cf}$ can
be close to the circular velocity.
For relatively condense preplanetesimals, their collision can eject material
into a satellite-forming disc. The disc can capture 
dust and boulders that enter the Hill sphere
after the encounter of RPPs.

One or several satellites of the primary could form in
this cloud (similar to typical models of formation of satellites of planets; 
see Woolson 2004, for example). The eccentricities
of such satellites would usually be small. As we show
at the end of Section 3.3, this scenario allows the formation
of satellites at practically any distance from the primary
up to the Hill radius. The initial radius of the `parent' preplanetesimal,
however, could be greater than the separation 
between components of the formed binary.

\subsection{Hybrid scenario and the formation of elongated bodies}

The two scenarios presented in Sections 2.2 and 2.3 could take place
at the same time. In addition to massive primary
and secondary components, smaller satellites moving
around either body could be formed.
For binaries formed in such a way, the separation distance
between the main components can be large or small.

It is possible that massive yet rarified binary components
merge when their central parts became dense enough. In
such a case, the form of the solid body obtained could differ
greatly from an ellipsoid. For example, this situation could give
rise to an elongated, bone-like body where both of the original
components are visible. As in other scenarios, the solid body
may have one or more small satellites. 
For example, (216)
Kleopatra (with dimensions of $217\times 94\times 81$ km) could
have formed in this manner. 

Even more rarely, several partly compressed components could merge
simultaneously.  In this case the form of the
solid body would be complicated, to some extent
`remembering' the forms of its components.

\section{THE ANGULAR MOMENTUM OF COLLIDED RAREFIED PREPLANETESIMALS}

\subsection{Basic model of an encounter between preplanetesimals}

In order to analyse the scenarios discussed in Section 2, 
Sections 3.2--3.4 will study the angular momentum $K_s$ of two
colliding RPPs relative to their centre of
mass and the period $T_s$ of the solid primary formed by contraction. 
Maximum and typical values of the separation distance
$L$ between binary components for the above scenarios
are discussed in Section 3.3.

Most of the formulae and calculations below are presented
for a `basic' model: the merger of two spherical RPPs
that before the collision were moving in coplanar, circular, heliocentric
orbits. The mass of the resulting object is equal to the sum of the two
preplanetesimal masses. The rotating, merged system
then contracts and tranforms into a solid planetesimal.
The angular momenta presented in Table 1 were obtained assuming that
the radii of the two preplanetesimals are equal to their Hill
radii. 
Based on the 'basic' model, we discuss more complicated models.

Scientists who study the formation of preplanetesimals generally
consider almost circular orbits and very small inclinations. These
conditions minimise the relative velocities of neighbouring preplanetesimals.
The probability of a merger is smaller for greater eccentricities. 
In all existing models of planetesimal discs, 
the mean inclination (in radians) of the orbits is assumed to be smaller than the mean
eccentricity. Relaxing this condition on the inclinations will not change
the order of magnitude of the angular momentum of two colliding
preplanetesimals, but can change its direction (see Sections 4.3 and 5.3).

For formation of a binary, it is not necessary that every
encounter between preplanetesimals (i.e., when one object's
centre of mass passes through the Hill sphere of the other) results
in a merger. It is enough that for some preplanetesimals,
at least one such encounter occurred during their lifetimes.
For almost circular 
heliocentric orbits, the minimum distance between 
encountered material points with masses of 
preplanetesimals can be several orders of magnitude smaller than 
the Hill radius $r_H$, 
and some of the material points can move inside the sphere for a long time.
For example, in an earlier work (Ipatov 1987) I presented a case where
two planetesimals move together for more than half an orbital period 
(see also discussions in Sections 2.1 and 5.1). Therefore, the values of $K_s$
and $T_s$ obtained for the `basic' model can also be true for 
encounters between preplanetesimals much smaller than their Hill
spheres. The sizes and mergers of encountered preplanetesimals
are discussed in Section 5.1.

Not all boulders could be captured during an encounter,
and some material could be lost at the stage of contraction.
Therefore, the mass of the formed binary (or a preplanetesimal)
may be smaller than the sum of the masses of the
two colliding preplanetesimals. Likewise, the final value of $K_s$ 
may be smaller than the total angular momentum of the RPPs.

Sections 3.2 and 3.3 present formulae for $T_s$ and $K_s$ 
after a single collision in the basic model. For multiple collisions, 
the angular momenta are summed. In my opinion (see
Section 4.3), most of the angular momentum possessed by a 
trans-Neptunian binary usually resulted from a single collision.

Naturally, a present-day binary can differ significantly from the primordial
binary formed by contracting RPPs. For example, tidal forces can increase the separation 
between binary components (e.g. the Pluto-Charon and
Antiope-S/2000(90) systems) and decrease the period of
axial rotation (as occurred for the Earth-Moon system) on
time scales shorter than the age of the Solar system. 
Although a given primordial
binary system might have been more compact than its 
corresponding discovered system, the total angular momentum 
could remain the same. The angular momenta of individual components 
could also change due to collisions with other small bodies and
the YORP effect. Finally, the angular momentum of a contracting preplanetesimal
could decrease due to tidal interactions with the Sun.

It is likely that most primordial planetesimals were 
ellipsoidal. 
The velocities of particles  close to the equator 
of a rotating contracting preplanetesimal %
were 
greater than those of particles  close to the poles,
and more particles could leave the equator. %
Therefore, some contracting preplanetesimals could be elongated
along the axis of rotation. On the other hand, a surrounding disc
could feed material to the equator, resulting in an oblate shape. In either case, 
the equator would be approximately circular. In the
`rubble pile' model considered by Walsh et al. (2008),
the axial radius was smaller than the equatorial radius. The formation
of highly elongated primordial small bodies was discussed
in Section 2.4.

\subsection{Principal formulae for calculation of the angular
momentum of two colliding preplanetesimals}

Previous papers (e.g. Lissauer et al. 1997; 
Ohtsuki \& Ida 1998) devoted to the formation of axial rotation of celestial bodies 
have mainly studied the final rotation rates of planets accreted from a disc
of solid planetesimals. My own studies on the formation of axial
rotation of rarefied preplanetesimals, solid planetesimals
and planets were presented in two preprints (N 101 and N
102) published in Russian in 1980 by the Institute of Applied
Mathematics of the USSR Academy of Sciences. These
studies are summarised in Ipatov (2000), which cites many
papers devoted to the formation of spin. The preprints
focussed mainly on the axial rotation of planets, but the formulae
used are applicable to a wide variety of cases. Sections
3.2--3.4 present similar formulae. 
 In Sections 4--5, %
the formulae are applied to the studies of the formation of binaries. %

Consider an inelastic collision between two non-rotating
spherical objects (preplanetesimals or planetesimals) with
masses $m_1$ and $m_2$ and radii $r_1$ and $r_2$, moving around the
Sun in coplanar circular orbits. The first object
is closer to the Sun. Its semimajor axis is denoted by $a$;
its circular velocity can be calculated from
the relationship $v_1 \equiv v_c=(G \, M_S/a)^{1/2}$,
where $G$ is the
gravitational constant and $M_S$ is the mass of the Sun. The
angular momentum of the colliding system relative to its
centre of mass is
\begin{equation}
K_s=v_{\tau} (r_1+r_2)  m_1  m_2/(m_1+m_2), 
\end{equation}
where $v_\tau$ is the tangential component of the velocity 
(with respect to the centre of mass of the first object)
at which the objects collide. 
If the difference $\Delta a$ in the orbits'
semimajor axes equals $\Theta (r_1+r_2)$, we have
\[
v_{\tau}=k_{\Theta}  v_c (r_1+r_2)/a
= k_{\Theta} (G \, M_S)^{1/2}  (r_1+r_2)  a^{-3/2}
\]
and 
\begin{equation}
K_s \equiv K_{s\Theta}=k_{\Theta} (G \, M_S)^{1/2} 
(r_1+r_2)^2 
\frac{m_1 m_2}{(m_1+m_2)a^{3/2}}. 
\end{equation}
For $r_a=(r_1+r_2)/a \ll \Theta$, 
one can obtain 
\begin{equation}
k_{\Theta} \approx (1-1.5 \Theta^2) 
\end{equation}
and
\begin{equation}
v_{rel}\approx(1-0.75\Theta^2)^{1/2} v_c r_a ,
\end{equation}
where $v_{rel}$ is the absolute value of the relative velocity between
the two colliding objects.

Below I derive equations (3) and (4). The angle between the lines 
connecting the Sun to the two encountering objects is 
$\alpha \approx \sin \alpha \approx r_a (1-\Theta^2)^{1/2}$.
 Consider the triangle ABC, where points A and B are
the positions of the objects and point C is the
intersection between the orbit of the first (faster) object
and the line from the Sun to the second object (point
B). The angle CAB is close to the angle $\gamma$  between the 
velocity of the first object and the direction from the
first object to the second object (located farther from the
Sun). We therefore consider $\sin \gamma \approx \Theta$.
Taking into account that  $v_c \propto a^{-1/2}$
and $(1+r_a \Theta)^{-1/2} \approx 1-r_a \Theta / 2$,
we obtain that the velocity of the second object equals
$v_2 \approx v_1 (1 -  r_a \Theta / 2)$. 
For the considered model,
$\cos \alpha \approx 1$ and $\cos \gamma \approx (1 - \Theta^2)^{1/2}$. 
We therefore obtain
$\sin (\gamma + \alpha) \approx \sin \gamma + \alpha \cos \gamma$,
and the difference between the tangential  velocities
of the two encountering objects is
\[
v_{\tau 1} - v_{\tau 2} = v_1 \sin \gamma - v_2 \sin (\gamma + \alpha)
\]
\[
\approx  v_1 \Theta - v_1 (1 -r_a \Theta /2) 
(\Theta + (r_a (1 - \Theta^2)^{1/2}) 
(1 - \Theta^2)^{1/2}) 
\]
\[
\approx v_1 r_a (1.5 \Theta^2 - 1) .
\]
This difference is valid for clockwise motions (negative angular
velocities). Therefore,  $k_\Theta \approx (1 - 1.5 \Theta^2)$.

Let us take into account that the angle between
velocities $v_1$ and $v_2$ is 
equal to $\alpha$, $v_2 \approx v_1 (1 -  r_a \Theta / 2)$,
$\sin \alpha \approx r_a (1- \Theta^2)^{1/2}$, and 
$\cos \alpha \approx 1$. Neglecting the term with $r_a^4$, 
we obtain  

\[
v_{rel}^2 \approx (v_1 - v_2 \cos \alpha)^2 + (v_2 \sin \alpha)^2 
\]

\[
\approx 
(\Theta^2/4 + (1 - \Theta^2)) (v_c r_a)^2 = (1 - 0.75 \Theta^2) (v_c r_a)^2 .
\]

Though the above derivation of equation (3) is not valid
for small $\Theta$,
Below we can use it to estimate the mean values of $k_{\Theta}$, $K_s$, and $v_{\tau}$,
 by evaluating the integrals over $\Theta$. 
 The reviewer noted that 
equations (3) and (4) are actually accurate even for $\Theta=0$, 
as planetesimals `colliding` (or more accurately, touching) 
on the same circular heliocentric orbit 
have nonzero angular momentum relative to each 
other in the inertial frame despite being stationary in the rotating frame. 
Only at larger $\Theta$ 
does the Keplerian shear begin to dominate,
resulting in a negative (i.e. retrograde) total angular momentum.


 The overall mean value of $\mid k_{\Theta}\mid$ is 0.6. Among positive values
of $k_{\Theta}$ the mean is $\frac{2}{3}$, and among negative values it is
$-0.24$. The values of $K_{s\Theta}$ and $k_{\Theta}$ are positive in the range
$0<\Theta<{\frac{2}{3}}^{1/2}\approx0.8165$ and negative in the range 
$0.8165<\Theta<1$.  
The same ranges of $\Theta$ for positive and negative values of
$K_{s\Theta}$ were used by Eneev Kozlov (1982), who obtained
them from a numerical integration of the
three-body problem. The minimum value of $k_{\Theta}$ is $-0.5$, and
its maximum value is 1. If $\Theta$ is uniformly distributed 
for objects moving in the same plane, the probability of 
a single collision producing negative rotation is about $\frac{1}{5}$.  
The ratio of the sum of positive values of $K_s$  to the
sum of all $\mid K_s \mid$ equals 0.925. This ratio, obtained
by evaluating the integrals given above, is close to the ratio obtained in
numerical experiments with initially circular orbits. 
The results of the experiments were presented
in the aforementioned preprints in 1981.

In the basic model, a new object forms in any encounter, and 
heliocentric orbits are considered to be circular at the time of encounter up to $r_H$.
This model could work for some encounters of very rarefied preplanetesimals
(see discussion in Section 5.1).
 Compared to the above model, 
 the fraction of collisions with positive angular momentum would be smaller
if one considers collisions of solid (or at least higher-density) objects 
and/or supposes that their orbits are circular only so long as the objects
are widely separated (that is, the orbits deviate from a heliocentric circle
when separation equals to $r_H$).
 Under the right conditions, 
it is even possible for 
the merged planetesimal to have negative momentum. 
Therefore, the dominance of positive angular momentum in
the basic model does not contradict the negative spins obtained
by authors who have used other
models and considered collisions of solid bodies (e.g. Giuli 1968).

Below we assume that a new spherical object of radius $r_{f\circ}$ and 
mass $m_f=m_1+m_2$ forms as the result of a collision, and then contracts to radius $r_f=r_{f\circ}/k_r$. If the angular momentum 
$K_{s\Theta}$ relative to the centre of mass 
is the same 
as that of the system at the time of collision, 
then equation (2) yields 
the period of axial rotation: 
\begin{equation}
T_s= \frac{2 \pi J_s}{K_{s}} =
k_T \frac{(m_1+m_2)^2  r_f^2 a^{3/2}}{m_1  m_2 (r_1+r_2)^2} ,
\end{equation}
where $k_T = 0.8 \pi \, \chi \, k_{\Theta}^{-1} 
(G \, M_S)^{-1/2}$. The moment of inertia is
$J_s=0.4 \chi m_f  r_f^2$,
where $\chi=1$ for a homogeneous sphere.
At $r_{f\circ}^3=r_1^3+r_2^3$, $m_1=m_2$, and $r_1=r_2$,
using equations (1) and (5), we then obtain
\begin{equation}
T_s=  2^{2/3}  k_T a^{3/2} k_r^{-2} \approx
{1.6\pi  \chi  r_f^2}  / (v_{\tau} r_1) 
\approx 6.33 \chi \, r_f / (v_{\tau} k_r) . 
\end{equation}
For $m_1\gg m_2$ and $r_f \approx r_1/k_r$, we have
\begin{equation}
T_s \approx k_T  (m_1/m_2)  a^{3/2}  k_r^{-2}
\approx 0.8\pi \, \chi \,  r_1 m_1 / (m_2 v_{\tau} k_r^2) .
\end{equation}
In the case of heliocentric orbits 
with eccentricity $e$, we can use 
$v_{\tau} \approx  2^{-1/2} \, v_c \, e $.

The angular momentum of binary components (with masses $m_p$ and $m_s$
 and radii $r_p$ and $r_s$) relative to their centre of mass equals 
\begin{equation}
K_{scm}=v_b  L \, m_p m_s/(m_p+m_s), 
\end{equation}
where $L$ is the 
mean distance between the components (i.e. the semimajor axis of their mutual orbit) 
and $v_b=(G(m_p+m_s)/L)^{1/2}$ is the mean velocity of their 
relative motion. The ratio 
of the momentum $K_{s\Theta}$ obtained at 
the collision of two spheres -- preplanetesimas 
to  $K_{scm}$ is
\begin{equation}
k_K = \frac{K_{s\Theta}}{K_{scm}} =
\frac {k_\Theta (M_S)^{1/2} (m_p+m_s)^{1/2} m_1  m_2 
(r_1+r_2)^2} {m_p  m_s (m_1 + m_2)  a^{3/2} L^{1/2}} .
\end{equation}

\subsection{The spins of colliding preplanetesimals that
have radii proportional to their Hill radii}

This subsection calculates $K_{s\Theta}$ under the assumption that
the radius of each preplanetesimal is proportional to the radius
of its Hill sphere, i.e. $r_{i}= k_H a (m_i/3M_S)^{1/3}$ ($i=1,2$). The coefficient
$k_H$ is arbitrary; as discussed in Section 3.1, we often
consider $k_H=1$. The radii used to calculate $K_{s\Theta}$ can be
much larger than the physical radii of the preplanetesimals.

Let us consider 
 $m_1=m_2=m_p=m_s=m_f/2=\frac{4}{3} \pi  r^3 \rho $ 
 and $r_p=r_s=r$, 
where $\rho$ is the density of solid binary components of radius $r$.
In this case, formulae (2), (8), and (9) yield 
\begin{equation}
K_{s\Theta} = 6^{-2/3} k_H^2 k_{\Theta} M_S^{-1/6} G^{1/2} a^{1/2} m_f^{5/3}  
\end{equation}
and
\begin{equation}
k_K^2= 6M_S k_{\Theta}^2 k_r^4 r / \pi \,\rho \, L \, a^3
\approx 3 \, M_S^{-1/3} a\, r \, L^{-1}  \rho^{1/3}  k_{\Theta}^2  k_H^4 .
\end{equation}
From equation (11) we obtain 
\[
L\approx 1.5 k_{\Theta}^2  k_H^4 k_K^{-2} r_H ,
\]
where $r_H \approx a (m_f/3M_S)^{1/3}$.
For $k_K = k_H =1$, we have $L \approx 0.5 r_H$ 
at $k_{\Theta}=0.6$,
and $L \approx  r_H$ at $k_{\Theta}=0.816$.
Therefore, for an
encounter between rarefied preplanetesimals moving in
almost circular heliocentric orbits, a binary with components 
of equal mass could form with any separation up to the
Hill radius.

In Sections 4.2–-4.3, I will compare the values of 
$K_{s\Theta}$ and $K_{scm}$ for several discovered binaries. Using the dependence
(11) of $k_K$ on $\rho$ and $L$, one can estimate the uncertainty on $k_K$ for
the binaries where these values are not well known.

At $r_{f\circ}= k_{Hpr}  r_H$, 
the ratio $k_r=r_{f\circ}/r_f$ 
is proportional to $a \, \rho^{1/3}$ and does not depend on mass. 
It equals 133 at $a =1$ au, $k_{Hpr}=1$, and $\rho=1$ g cm$^{-3}$.
For a solid planetesimal 
formed by contraction of a preplanetesimal of radius $r_{f\circ}$, 
one obtains
\[
T_s \propto a^{3/2}  k_r^{-2} \propto a^{-1/2}  \rho^{-2/3}. 
\]

At $\chi=1$, $k_\Theta=0.6$, $a$=1 au, $k_{H}=k_{Hpr}=1$, and
$m_1=m_2$, the rotation period of a RPP
formed from the collision of two Hill sphere-sized
preplanetesimals is $T_s\approx 9\cdot 10^3$ h. At $k_r$=133, 
on the other hand, $T_s\approx 0.5$ h.
For greater values of $a$, $T_s$ is even smaller. Such small periods
of axial rotation cannot exist, especially as
bodies obtained by the contraction of rotating rarefied preplanetesimals
are loosely bound, and can lose material easier than solid bodies.

The value of $v_s=2\pi \, r_f/T_s$ 
(the velocity of a particle on the surface of a rotating spherical object of 
radius $r_f$ at the equator) is equal to $v_{cf}=(G \, m_f/r_f)^{1/2}$ (the 
minimum velocity of a particle that can leave the surface) at 
$T_{sc}=T_s=(3\pi / \rho G)^{1/2}$. Thus, $v_s$ is equal to the equatorial escape velocity $v_{cf}$
at $T_{se}=T_s=(3\pi / 2\rho G)^{1/2}$. If $k_{\rho}$ is the 
density in g cm$^{-3}$, then $T_{sc} \approx 3.3/k_{\rho}^{1/2}$~h and 
$T_{se} \approx 2.33/k_{\rho}^{1/2}$~h.

Considering that $T_{sc} = 2 \pi J_s /K_{s\Theta}$, the critical radius
at which material begins to leave a contracting preplanetesimal 
is equal to
\begin{equation}
r_{cr}= \frac {K_{s\Theta}^2} {0.16 \chi^2 m_f^3 G}
= \frac {k_{cr} k_{\Theta}^2 k_H^4 r_H }{\chi^2}.
\end{equation}
The coefficient $k_{cr}$ in equation (12) is approximately 0.82 at $m_1=m_2=m_f/2$,
and $k_{cr} \approx 2.08 (m_2/m_1)^2$ at $m_2 \ll m_1 \approx m_f$.
Therefore, at $k_{\Theta}= k_H= \chi =1$ and $m_1=m_2$,
the radius of the disc formed around the primary
could be as much as $0.82r_H$.
 In most cases, however, the disc would be relatively small 
since the masses of colliding preplanetesimals are usually different
and the preplanetesimals are smaller than their Hill spheres.

If preplanetesimals are smaller than their Hill spheres, 
it may be better to consider the encounter of the 
preplanetesimals up to the radius 
of the Hill sphere corresponding to the mass 
$m_f=m_1+m_2$,
but not up to the sum of the radii of their Hill spheres,
as it was in formulae (2), (5) and (10). 
For two identical preplanetesimals, this consideration
decreases $K_{s\Theta}$  and increases $T_s$ by a factor of 4$^{2/3} \approx 2.5$.

\subsection{The spin imparted to a preplanetesimal by the accretion
of much smaller objects}

A preplanetesimal can change its spin and mass due
to the capture of much smaller objects (e.g. dust and boulders)
initially moved outside of its Hill sphere. 
In this subsection,
we consider only accreted objects initially moving in circular heliocentric orbits.
At $\Delta a=\Theta \, r_{12}$, the difference in 
the velocities of two objects is about 
$0.5 v_c \Theta r_{12} / a$, where $r_{12}=r_1+r_2$. 
The relative distance covered with this velocity during one revolution
around the Sun is $(0.5 \Theta v_c r_{12} /a) (2 \pi a/ v_c) = \pi \Theta  r_{12}$. 
Thus, about $(2 \pi a) / (\pi \Theta r_1) = 2a / 
(\Theta \, r_1)$ 
revolutions are needed for an object with radius $r_1$ to sweep up all
smaller objects with radii $r_2 \ll r_1$.
For example, taking 
Pluto's Hill sphere and semimajor axis for $r_1$ and $a$, 
we obtain $2a/ r_1 \approx 1.5 \cdot 10^4$. This value may be 
similar to the contraction times of trans-Neptunian preplanetesimals, or smaller
by one or two orders of magnitude (see the discussion in Section 1.1).

A cross-section of a sphere is a circle, and the length of a chord 
at distance $\Theta \,r_1$ from its centre is 
$2 r_1 (1-\Theta^2)^{1/2}$. The relative velocity at this distance
is proportional to $\Theta r_1 / a$. Therefore, 
if small objects are uniformly distributed  around the orbit
of the first object, 
the number of the small objects captured by a preplanetesimal 
per unit of time
at $\Delta a=\Theta \,r_1$  is proportional to 
$(1-\Theta^2)^{1/2} \Theta $. Multiplying this 
value by $k_{\Theta}$ (as $K_s \propto k_{\Theta}$), we obtain 
$K_s \propto \Theta (1-\Theta^2)^{1/2}  (1-1.5 \Theta^2$). 
 By integrating over
$\Theta$ from 0 to 0.8165 and from 0.8165 to 1 for a circular cross-section, 
I find that the ratio of positive to negative angular momenta is 9.4.

Let us consider a spherical preplanetesimal of mass
$m_{pp}$ and radius $r=r_{pp}$ that grows due to collisions 
with smaller objects as described above. 
Typical tangential velocity of collisions is supposed to be
$v_{\tau}=0.6 v_c  r/a$, and $\Delta K$ is the difference between 
fractions of positive and negative angular
momenta acquired by the preplanetesimal.
The ratio between the radii of a 
preplanetesimal of density $\rho_\circ$ and the planetesimal
formed by its contraction is denoted by
$k_r$.
Considering an integral over a radius $r$ of a growing preplanetesimal
and supposing $dK_s = r \, v_{\tau} dm$ and $dm=4 \pi \rho_{\circ} r^2 dr$,
we can calculate $K_s$. 

Finally, the rotation period of the
planetesimal formed by contraction of the preplanetesimal is
\[
T_s= 2 \pi J_s / K_s \approx 7  \chi  \, a^{3/2}  (G \, M_S)^{-1/2}  \Delta K^{-1}  k_r^{-2} .
\]
If $r_{pp}$ is equal to the radius of the Hill sphere, we have
\[
T_s=5.7 \chi \, M_S^{1/6}  (G \, a)^{-1/2} \Delta K^{-1}
\rho_{\circ}^{-2/3} k_r^{-2}. 
\]
If the preplanetesimal is much smaller than its Hill sphere and
$v_{\tau}=\alpha \, v_{par} $ (where $v_{par}$ is the parabolic
velocity at the surface of the preplanetesimal), then 
\[
K_s \approx 0.85 \alpha (r_{pp} G)^{1/2} m_{pp}^{3/2} \Delta K
\approx 0.67 \alpha G^{1/2} \rho_{\circ}^{-1/6} m_{pp}^{5/3} \Delta K
\] and
\[
T_s \approx 1.45 \chi (\alpha  \Delta K)^{-1} (\rho_{\circ}  G)^{-1/2} k_r^{-2}. 
\]

Based on simulations of a large number of encounters between
two objects of density $\rho$ moving around the Sun, we calculated
values of $\Delta K$ for different elements of heliocentric orbits.
The calculations showed that the fraction $K_{+}$ of positive angular momenta (note that $\Delta K=2K_{+}-1$) tends to decrease with increases in $a$, $\rho$, or eccentricity.
 For RPPs and small eccentricities $e$, positive
angular momentum increments are more frequent than
negative increments ($K_{+}>0.9$ at $e$=0). 
For solid objects in orbits with $a=1$ au and $e\cdot (m_1/M_S)^{-1/3} \sim 2-7$,
$\Delta K$ is positive and about a few hundredths.
For circular orbits, $\alpha$ was typically calculated to be 0.6.

\section{COMPARISON OF ANGULAR MOMENTA OF DISCOVERED BINARIES WITH MODEL ANGULAR MOMENTA}

\subsection{Description of the data presented in Table 1}

For six discovered binaries, 
the angular momenta $K_{scm}$ of the systems
were estimated and compared to the present model. 
The primary and secondary 
components of the binary systems have diameters
$d_p$ and $d_s$ and masses $m_p$ and $m_s$, reported in
the first rows of Table 1. Using the formulae presented in 
Sections 3.2 and 3.3, I also calculate the theoretical 
angular momentum $K_{s06ps}$ of 
two preplanetesimals with  masses $m_p$ and $m_s$
 encountered from 
circular heliocentric orbits at the mean value of $k_\Theta$ 
(equal to 0.6), and the angular momentum 
$K_{s06eq}$ of two preplanetesimals with 
masses equal to 0.5($m_p+m_s$).       
Thus, each system is compared to a theoretical scenario
where two rarefied preplanetesimals on different orbits merged then
contracted into a pair of solid bodies. All three momenta are relative
to the centre of mass of the system. 

While $K_{s06ps}$ and $K_{s06eq}$ are calculated for RPP
diameters equal to the Hill radius, 
binary systems can also be formed by the collision of much smaller 
preplanetesimals, as discussed in Section 3.3. Naturally, the 
heliocentric orbits of
two preplanetesimals can have other separations, so
$k_\Theta$ can be smaller or larger than 0.6.

The spin momentum of the primary is 
$K_{spin}=0.2 \pi \, \chi \, m_p  d_p^2 /T_{sp}$,
where $T_{sp}$ is the period of axial rotation of the primary. 
The separation distance between the primary and the secondary is denoted by $L$.
The data used in these calculations were taken from the site
http://www.johnstonsarchive.net/astro, which also provides 
many references for each system.
Similar (but older) data can be found in Richardson \& Walsh (2006). For calculation of 
$r_{Kps}=(K_{scm}+K_{spin})/K_{s06ps}$ and 
$r_{Keq}=(K_{scm}+K_{spin})/K_{s06eq}$, 
only the spin of the primary was considered. However, these
ratios would be about the same if they included the spins of all 
components, because $K_{spin}$ is several times smaller 
than $K_{scm}$ even for equal masses. 
For unequal masses most of the spin  momentum 
is due to the primary component. 

\begin{table*}
\begin{minipage}{160mm}
\caption{Angular momenta of several binaries.} \label{anymode}
\begin{tabular}{@{}lllllll}
\hline
binary & Pluto & (90842) Orcus  & 2000 CF$_{105}$ & 2001 QW$_{322}$ & (87) 
Sylvia & (90) Antiope \\
\hline
$a$, au          & 39.48  & 39.3  &  43.8  & 43.94  & 3.94 &  3.156 \\
$d_p$, km        & 2340   & 950   &  170   & 108?   & 286  &  88    \\
$d_s$, km        & 1212   & 260   &  120   & 108?   & 18   &  84    \\
$m_p$, kg  & $1.3\times10^{22}$ & $7.5\times10^{20}$ & $2.6\times10^{18} $ ?&
 $6.5\times10^{17}$ ?   & $1.478\times10^{19}$   &  $4.5\times10^{17}$    \\
$m_s$, kg  & $1.52\times10^{21}$ & $1.4\times10^{19}$  & 
$9\times10^{17} $ ?& $6.5\times10^{17}$ ?  & $3\times10^{15}$ &  $3.8\times10^{17}$    \\
  &   &   for $\rho$=1.5 g cm$^{-3}$ & &  &  for $\rho$=1 g cm$^{-3}$  &  \\
$L$, km        & 19,750 & 8700 & 23,000  & 120,000 & 1356  & 171 \\
$L/r_{H}$  & 0.0025   & 0.0029  &  0.04  & 0.3   & 0.019  &  0.007 \\
$2L/d_p$  & 16.9  & 18.3  &  271  & 2200   & 9.5  &  3.9 \\
$T_{sp}$, h  & 153.3   & 10  &     &    & 5.18  &  16.5 \\
$K_{scm}$, kg~km$^2$~s$^{-1}$ & $6\times10^{24}$ & $9\times10^{21}$ &
$5\times10^{19}$ & $3.3\times10^{19}$ & $10^{17}$ & $6.4\times10^{17}$ \\
$K_{spin}$, kg~km$^2$~s$^{-1}$ & $10^{23}$ & $10^{22}$ &
$1.6\times10^{18}$ & $2\times10^{17}$ & $4\times10^{19}$ & $3.6\times10^{16}$ \\
 & & & at $T_s$=8 h& at $T_s$=8 h & &  \\
$K_{s06ps}$, kg~km$^2$~s$^{-1}$ & $8.4\times10^{25}$ & $9\times10^{22}$ &
$1.5\times10^{20}$ & $5.2\times10^{19}$ & $3\times10^{17}$ & $6.6\times10^{18}$ \\
$K_{s06eq}$, kg~km$^2$~s$^{-1}$ & $2.8\times10^{26}$ & $2\times10^{24}$ &
$2.7\times10^{20}$ & $5.2\times10^{19}$ & $8\times10^{20}$ & $6.6\times10^{18}$ \\
$(K_{scm}+K_{spin})/K_{s06ps}$  & 0.07   & 0.2  &  0.3 & 0.63   & 130  &  0.1 \\
$(K_{scm}+K_{spin})/K_{s06eq}$  & 0.02   & 0.01  &  0.2 & 0.63   & 0.05  &  0.1 \\
$v_{\tau eq06}$, m~s$^{-1}$ & 6.1   & 2.2  &  0.36 & 0.26   & 2.0  &  0.82 \\
$v_{\tau pr06}$, m~s$^{-1}$ & 5.5   & 1.8  &  0.3 & 0.26   & 1.3  &  0.82 \\
$v_{esc-pr}$, m~s$^{-1}$ & 15.0   & 5.8  &  0.8 & 0.53   & 5.3  &  1.7 \\
\hline
\end{tabular}
\end{minipage}
\end{table*}

Table 1 also reports the values of $2L/d_p$ and
$L/r_H$, where $r_H$ is the Hill radius for the total
mass $m_{ps}$ of the binary. Three velocities are presented in
the last lines of Table 1: $v_{\tau pr06}$ is the tangential velocity
$v_\tau$ of the 
encounter at $k_\Theta=0.6$ between Hill spheres 
with masses equivalent to the present primary and secondary components; 
$v_{\tau eq06}$ is the same velocity for Hill spheres of equal masses (0.5$m_{ps}$);
and $v_{esc-pr}$ is the escape velocity at the edge of the
Hill sphere of the primary.

A relative velocity $v_{rel}$ equal to 1 m~s$^{-1}$ is obtained
for two colliding objects orbiting at $e\sim0.0002$ and $a$=40 au.
The model described above uses perfectly circular heliocentric orbits, 
and may not be correct at $e>0.0002$ for preplanetesimals 
corresponding to solid objects of diameter $d\le200$ km because
$v_{rel}\sim v_{esc-pr}$ for $e\sim 0.0002$ and $d\sim 200$ km. 
This conclusion is valid for other values of $a$, since both $v_{rel}$ and 
$v_{esc-pr}$ are proportional to $a^{-1/2}$. It is reasonable to assume 
that the eccentricities of rarefied preplanetesimals are very small because of
their interaction with gas and dust.

\subsection{Discussion of the data presented in Table 1}

For the trans-Neptunian binaries presented in Table 1, 
neither $r_{Kps}=(K_{scm}+K_{spin})/K_{s06ps}$
nor $r_{Keq}=(K_{scm}+K_{spin})/K_{s06eq}$ 
($r_{Kps}>r_{Keq}$) exceed 0.63. This highest value
was obtained for 2001 QW$_{322}$, which consists of two approximately
equal masses separated by a large distance. For the
other trans-Neptunian binaries, $L/r_H$ 
and, therefore, $r_{Keq}$ 
are smaller (often
much smaller) than those for 2001 QW$_{322}$. 
The small values of $r_{Kps}$ and $r_{Keq}$
can be explained if the discovered binaries arose from preplanetesimals
that were already partly compressed (i.e. smaller than their Hill spheres) 
and/or denser towards the centre at the moment of collision. 
Petit et al. (2008) noted that most 
models of binary formation cannot easily explain
the properties of 2001 QW$_{322}$. 
For this binary,
the equation $K_{s\Theta}=K_{scm}$ is fulfilled if 
$k_\Theta\approx 0.4$ and $v_\tau \approx 0.16$ m~s$^{-1}$. 
Therefore,  formation of this binary system can be
explained as the merger of two RPPs in circular
heliocentric orbits.  

Another 
property of 2001 QW$_{322}$ is that its 
components have a retrograde mutual orbit.
As shown in Section 3.2, the angular momentum is 
negative for encounters of RPPs at $0.82<\Theta<1$.
This condition can also result in the formation of widely separated binaries.

Main-belt asteroid (87) Sylvia (384$\times$262$\times$232 km) has
two satellites, Romulus  ($L_1$$\approx$1356 km, 
$d_1$$\approx$18 km) and Remus  ($L_2$$\approx$706 km,
$d_2$$\approx$7 km).  For this system, $r_{Keq} \approx 0.05 \ll r_{Kps} \approx 130$. 
If the Sylvia system originated in
the collision of two rarefied preplanetesimals, then their masses 
did not differ by more than an order of magnitude (as $K_{s06eq}/K_{spin} \approx 20$).

Main-belt asteroid (90) Antiope belongs to a synchronous
double-asteroid system. Its rotational and orbital periods 
are 16.5 h, and the mutual orbit of the components is
almost circular. Before tidal forces induced these orbits, 
the components may have been separated by a distance even smaller than the 
present 171 km. Since
the present ratio $2L/d_p$ is small, this binary likely
formed at the stage of solid (or almost solid) bodies. For this
binary, $r_{Kps} \approx r_{Keq} \approx 0.1$.  
Since $K_{scm}/K_{s\Theta}\propto (L/a)^{1/2}$ (see
Section 3.3), the model of rarefied Hill sphere preplanetesimals
in almost circular heliocentric orbits is capable of producing a binary system 
with masses of components similar to those for Antiope.
However, the separation $L$ obtained is 
greater by two orders of magnitude than that of the Antiope binary  
and is comparable to the radius of the Hill sphere.

\subsection{Formation of Pluto's axial rotation and the
inclined mutual orbits of binary systems}

Pluto has three satellites, but the contribution of two satellites (other than 
Charon) to the total angular momentum of the system is small; for the whole
system $K_{scm}/K_{spin}\approx60$. The axial tilt of Pluto with respect to its orbit is 119.6$^\circ$. 
Such a 
reverse rotation is possible in the collision of two (solid or rarefied) objects, but
at the stage of rarefied preplanetesimals, it is better to use 
$\mid k_{\Theta} \mid \; \sim 0.2-0.3$  (see
Section 3.2). In the proposed model, it is not possible to obtain Pluto's 
or any reverse rotation from
a large number of collisions with smaller objects (as in Section 3.4).

For a collision of two identical Hill sphere preplanetesimals, 
the value of $K_{s\Theta}$ obtained for $\mid k_{\Theta} \mid \; = 0.3$
exceeds $K_{scm}$ for the Pluto system by a factor of 23. 
 If the masses of the parent Hill sphere RPPs are equal to the masses of 
Pluto and Charon, then 
$K_{s\Theta}/K_{scm} \approx 7$ at $\mid k_{\Theta} \mid \; = 0.3$. 
 Furthermore, the angular momentum
of colliding preplanetesimals in eccentric orbits can be greater
than that obtained for circular orbits. The above estimates 
imply that the radii of the preplanetesimals that gave rise to
the Pluto system were smaller by at least an order
of magnitude than their Hill radii.

Most of the angular momentum of the Pluto system could
have resulted from a single collision of preplanetesimals 
moving in different planes. To explain Pluto's axial tilt 
(about 120$^\circ$) and the inclined mutual orbit of 2001 QW$_{322}$ 
(124$^\circ$ from the ecliptic) in these terms, I first note that the 
thickness of the disc (i.e., the distances between the
centres of mass of the colliding preplanetesimals and the 
equatorial plane of the disc) was comparable
to or greater than the radii of the RPPs themselves. 
The inclined mutual orbits of
many trans-Neptunian binaries support the idea that the
momenta of such binaries mainly resulted from single collisions of preplanetesimals
rather than the accretion of smaller objects. (In the latter case, 
the primordial inclinations of mutual orbits to the ecliptic 
would be relatively small.)

In the models of binary formation considered here,
the spin vector of the primary preplanetesimal is 
almost perpendicular to the plane in which satellites of
the primary move (as is the case with Pluto). However, the
direction of this vector could vary over time due to collisions of the primary 
with solid bodies.

\section{DISCUSSION}

\subsection{Radii and mergers of rarefied preplanetesimals}

The radii of preplanetesimals need to be comparable to their 
Hill radii only if the resulting binary is to have a
separation $L$ comparable to its own Hill radius $r_H$. 
Excepting 2001 QW$_{322}$, all the binaries
presented in Table 1 and many other discovered binaries
have ratios $L/r_H$ below 0.04. Among the binaries considered 
by Richardson \& Walsh (2006) in their fig. 1, this ratio was less
than 0.3. RPPs much smaller
(by at least by an order of magnitude) than $r_H$ suffice to
explain the formation of binaries with $L/r_{H}<0.04$.  Generally
speaking, the required
radii of the rarefied preplanetesimals can be of the same order 
as the present distance $L$ between the binary components. However,
the secondary component could form much closer to the primary than the outer 
edge of the disc, and $L$ could increase due to tidal forces. This estimate
of the preplanetesimal size is therefore not an accurate predictor.

The values of $2L/d_p$ presented in Table 1 vary from 17 to
2200 for trans-Neptunian binaries. The trans-Neptunian
binaries considered by Richardson \& Walsh (2006) fall inside the same range, 
but asteroid binaries may have smaller values. Cuzzi et al. (2008) considered 
spherical, rarefied clumps of diameter $l=(1-5)\times10^4$ km and mass
equivalent to a body of unit density and radius 10--100 km, orbiting
at $a$=2.5 au. The diameters $l$ 
are greater than the separations of known asteroid binaries.
Rarefied pre-asteroids may therefore have solidified 
before colliding with other pre-asteroids.

The densities of rarefied preplanetesimals can be very
low, but their relative velocities $v_{rel}$ during an encounter
are also very small. In particular, the relative speed of the collision is 
typically smaller than the escape velocity $v_{esc-pr}$ at the Hill radius 
of the primary. In addition to $v_{esc-pr}$, Table 1 presents the tangential
component $v_{\tau pr06}$ of $v_{rel}$. In some models of the evolution of
Saturn's rings (R. Perrine, private communication, 2009) 
colliding objects form a new object if their impact speed is less 
than the mutual escape speed by a certain factor. If the 
greater encountered object is much
smaller than its Hill sphere, and if both heliocentric
orbits are almost circular before the collision, then the velocity 
of the collision $v_{col}\approx (v_{rel}^2+v_{par}^2)^{1/2}$ 
does not differ much from the parabolic velocity $v_{par}$ 
at the surface of the primary RPP (radius $r_{pc}$). Indeed,
$v_{par}$ is proportional to $r_{pc}^{-1/2}$. 
Therefore, encounters could result in a merger
(followed by formation of a binary) at any $r_{pc}<r_H$.

For some pairs of objects, the mutual orbit
can be more complicated than a parabola or ellipse. In
a coplanar model, decreasing $r_{pc}$ by a factor of 10 can reduce 
the fraction of encounters resulting in a collision by a factor less than 10. 
Greenzweig and Lissauer (1990) published plots showing the dependence of the 
closest approach $r_{min}$ between objects 
in initially circular heliocentric  
orbits separated by an initial distance $\Delta a$. Based on these figures,
the maximum value of $\Delta a$ is $0.47r_H$
for $r_{min}<0.1r_H$ and $0.16r_H$ for $r_{min}<0.01r_H$. 
In Greenzweig and Lissauer's  
study, the integration 
began at a distance greater than $r_{H}$ and
considered close encounters taking place at $\Delta a<3.5r_H$. 
Note that in the above example, 
while the densities of uniform spherical preplanetesimals
ranging from $r_H$ to $0.01r_H$ in size differ
by six orders of magnitude, the probability of collision
differs only by a factor of $\sim 20$ provided all motions occur
in the same plane. The difference in the probabilities
is much greater in a non-planar model.

Johansen et al. (2007) determined that the mean free
path of a boulder inside a cluster -- preplanetesimal is shorter
than the size of the cluster. This result supports the picture
of mergers between rarefied preplanetesimals.
Lyra et al. (2008) noted that the velocity dispersion of rarefied preplanetesimals
remains below 1 m s$^{-1}$ in most simulations, so
destructive collisions between boulders are avoided.

To illustrate the probability of a merger between two rarefied
preplanetesimals, consider the following simple model. There is
a spherical preplanetesimal of diameter $D_s$ and mass $M$, consisting
of $N$ identical boulders of diameter $d$. After contraction,
a solid, spherical planetesimal of diameter
$D$ is formed. The densities $\rho$ of the boulders and the
solid planetesimal are the same. A second preplanetesimal then passes through 
this cluster.  The ratio of the length of its path inside the first
preplanetesimal to $D_s$ is denoted as $k_s$. 
Because the relative motion of the centres of mass of 
the preplanetesimals can be complicated, 
$k_s$ can exceed 1.

The boulders belonging to both preplanetesimals
are considered to have identical diameters $d$, and the volume
swept by one boulder is $\pi \, d^2  k_s D_s$.  The ratio of this
volume to the volume $\frac{\pi}{6} D_s^3$  of the Hill sphere, divided by
$N=(D/d)^3$, gives the number of collisions 
$N_{col} = 6 k_s D^3 / (D_s^2 d)$.
Let us take
$D_s = 2 k_H a (M/3 M_S)^{1/3}$,
$\rho = k_{\rho}$ g cm$^{-3}$, and $a = k_a$ au.
In this case, we obtain
\begin{equation}
D/d \approx N_{col} \times 3 \cdot 10^3 \cdot k_a^2 k_{\rho}^{2/3} k_s^{-1} k_H^2  .
\end{equation}
For $D=1000$ km, $d=0.3$ m,  and $k_{\rho} = k_s = k_H = N_{col} =1 $,
equation (13) is fulfilled  at $a = 33$ au.
Using equation (13),
we find that $N_{col} \propto D \cdot k_H^{-2}$. This verifies the
intution that for small values of the ratio $k_H$ between a preplanetesimal's 
radius and its Hill radius, $N_{col}$ can be relatively large. $N_{col}$ also decreases
with $D$, but at $D=50$ km
and $k_H = 0.2$, $N_{col}$ is almost the same as for $D=1000$ km and $k_H = 1$.
This means that for most binaries in the Solar system with $D>50$ km,
their parent preplanetesimals should have had
$N_{col}\ge 1$ provided that $d \le 1$ m.
 Boulders (or dust particles) are more likely to be captured if they have
smaller diameters $d$. 
The probability of capture also increases for boulders
(particles) closer to the centres of the preplanetesimals,
if  the density of the preplanetesimal is higher
at a smaller distance from the centre. 
The relative 
velocities of the boulders in this model are usually smaller than the 
escape velocities of the preplanetesimals, so some collided boulders could 
remain inside the Hill sphere. 

At greater eccentricities, 
the mean collision velocity between preplanetesimals and the 
minimum distance of closest approach
 between material points corresponding to encountering preplanetesimals 
are greater, and 
the particles  remain inside the Hill sphere for less time.
Therefore,
the probability of a merger of preplanetesimals is smaller. However, 
the typical angular momentum of preplanetesimals 
encountered up to the Hill sphere is greater.

The fraction of trans-Neptunian binaries that formed during the preplanetesimal stage 
depends on 
the initial distribution of sizes, densities, and contraction times among preplanetesimals. These issues must wait on further studies of the formation and evolution of preplanetesimals.

\subsection{Binaries that originated at different distances from the Sun}

Given a primary of mass $m_p$ and a much smaller secondary, both
in circular heliocentric orbits, one can obtain 
$v_{\tau}/v_{esc-pr} \approx 0.66 k_{\Theta} (\frac{M_S}{m_p})^{1/3} / a$ 
(see designations and formulae in Sections 3.2 and 4.1). 
This ratio decreases with increasing $a$ and $m_p$.
Therefore, preplanetesimals are more likely to merge when the primary is
more massive and located farther from the Sun. 
 The total mass of all preplanetesimals and the ratio of the time needed for 
contraction of  preplanetesimals to the period of their rotation around 
the Sun  could be greater for 
the trans-Neptunian region than for the initial asteroid belt.
 (Several authors have arrived at the similar conclusion on the ratio 
 for dust condensations, e.g. Safronov 1969.)

The above factors could explain why a
larger fraction of binaries are found at greater distances from the Sun,
and why the typical mass ratio (secondary
to primary) is greater for TNOs than for asteroids. The
binary fractions in the minor planet population are about 2
per cent for main-belt asteroids, 22 per cent for cold classical
TNOs, and 5.5 per cent for all other TNOs (Noll 2006). Note that TNOs 
moving in eccentric
orbits (the third category) are thought to have formed near the
giant planets, closer to the Sun than classical TNOs (e.g. 
Ipatov 1987).

\subsection{Asteroid binaries}

In our opinion, some asteroid binaries with large primaries
could have formed at the preplanetesimal stage. The eccentricities
of satellite orbits around large asteroids (including Sylvia and Antiope)
are usually (but not always) relatively small. Under the proposed model of 
rarefied preplanetesimal mergers, small eccentricities could indicate that
the satellites formed from a disc around the primary. The plane of the disc
will be close to that defined by the line connecting two preplanetesimals
and their relative velocity vector. If the initial preplanetesimal orbits have large
eccentricities $e$ and/or inclinations $i$, the relative velocity 
depends on the angle between
the planes of heliocentric orbits. 
For small $e$ and $i$, the plane
of relative motion inside the Hill sphere
can differ significantly from the orbital planes.

If the thickness of the disc of preplanetesimals
is of the same order as the diameters of the largest Hill spheres, then
the proposed model of binary formation permits any inclination $i_m$ 
between the mutual orbit of a binary and the ecliptic. 
Some asteroid binaries have highly inclined mutual
orbits (e.g.  $i_m$=64$^\circ$ for Antiope and $i_m$=93$^\circ$ for Kalliope).
For (87) Sylvia, (107) Camilla, and (283) Emma, on the other hand, $i_m$ does
not exceed 3$^\circ$.
 High-velocity collisions between solid asteroids
may more often yield $i_m$ 
of the order of typical inclinations in the asteroid belt, as most of the debris 
will  have neither large nor very small inclinations.

The total angular momentum of two identical Hill sphere preplanetesimals
in circular heliocentric orbits often exceeds (In the case of (87) Sylvia, by
an order of magnitude) the critical value at which particles 
leave the surface of the solid primary. 

Solid bodies can also attain the 
critical angular momentum (corresponding to $T_s \le 3.3/k_{\rho}^{1/2}$ h; 
see the end of Section 3.3) by collisions, but eccentric (and/or inclined)
heliocentric orbits are required. 
Using equation
(6) with $m_1=m_2$, $v_\tau/\chi=3.5$ km~s$^{-1}$ 
(e.g. at a typical velocity $v_{col}$ of 
collisions in the asteroid belt equal to 5 km~s$^{-1}$,
$v_{\tau} \approx 0.7v_{col}$, and $\chi =1 $),
 and 
$r_f\approx 6600$~km, I obtain $T_s \approx 3.3$ h. Therefore, the critical 
angular momentum can also result from the collision of two identical solid
asteroids of any radius ($<$6600 km). 

For a given ratio $m_1/m_2$ and eccentricities
of heliocentric orbits of solid bodies, 
the probability of attaining the critical momentum during
a collision is greater for smaller values of $m_1$ and $a$ (because
$T_s\propto r_f/v_\tau$, and at smaller $a$ both orbital velocity 
and $v_\tau$ are larger).
For $v_{\tau}/\chi=3.5$ km~s$^{-1}$, $T_s=3.3$ h, and $r_1=100$ km; formula (7) then yields $m_2/m_1\approx 0.006$.

Solid bodies can be disrupted during a
collision, whether or not they reach the critical angular momentum. Most of the material
ejected in this fashion would leave the Hill
sphere, especially when the primary has a relatively small
mass. The spin periods of asteroid primaries with $d>50$ km are
often several times greater than 3 h. It may be that
their original rotation was faster.

Some present-day asteroids (especially those with $d<10$
km) may be debris from larger solid bodies. As discussed in
Section 1.2, binary asteroids in the near-Earth object population
formed at the stage of solid bodies. The spin
and shape of these bodies could have changed during the
evolution of the Solar system due to collisions with other
small solid bodies.

\subsection{The colours and total mass of TNOs}

In the proposed model of binary formation, two 
preplanetesimals originate at almost the same
distance from the Sun. This point agrees with the correlation
between the colours of primaries and secondaries obtained by
Benecchi et al. (2009) for trans-Neptunian binaries. In addition,
the material within the preplanetesimals could
have been mixed before the binary components formed.

The formation of classical TNOs from rarefied preplanetesimals could have taken place for 
a small total mass of preplanetesimals in the trans-Neptunian region, even given 
the present total mass of TNOs.
Models of TNO formation from solid planetesimals 
(e.g. Stern 1995; Davis \& Farinella 1997; Kenyon \& Luu 1998, 1999)
require a more massive
primordial belt and small ($\sim 0.001$) eccentricities
during the process of accumulation. However, the 
gravitational interactions between planetesimals during this stage
could have increased the eccentricities to values far
greater than those mentioned above (e.g. Ipatov 1988,
2007).  This increase testifies in favour of formation
of TNOs from rarefied preplanetesimals.

\section{CONCLUSIONS}

This analysis has shown that some trans-Neptunian objects could 
have acquired their axial momenta and/or satellites during a primordial 
stage as rarefied preplanetesimals. In this scenario, most rarefied pre-asteroids
could have solidified before colliding with other pre-asteroids.

Models of binary formation due to gravitational
interactions or collisions between objects within the Hill radius,
which have been studied by several authors for solid objects, could
be more effective for rarefied preplanetesimals. For example,
due to their almost circular heliocentric orbits, two nearby preplanetesimals 
remain inside the Hill radius longer. The centres of mass 
of two rarefied preplanetesimals may also be able to approach each
other more closely than solid planetesimals because the planetesimals
could have greater eccentricities of heliocentric orbits.

During a collision, rarefied preplanetesimals can have
higher densities closer to their centres. In this case, there could be two centres of
contraction inside the rotating preplanetesimal
formed as a result of the collision of two rarefied preplanetesimals. 
The result would be a binary with two roughly equal masses,
which could be separated by any distance up to the Hill radius. 
The eccentricity and inclination
of the secondary component's orbit around the 
primary component can have any value. The observed
separation distance can be of the order of 
the radius of a greater encountered preplanetesimal.

Some binaries could form
because the angular 
momentum of a binary that was obtained at the stage of rarefied preplanetesimals 
was greater than the angular momentum that can exist for solid bodies.  Material that left a contracted preplanetesimal 
formed as a result of a collision of two preplanetesimals  
could form a disc around the primary.
One or more satellites could  grow in the disc
within the Hill radius, typically much less. These satellites move mainly
in low-eccentricity orbits.

The two scenarios described above can take place at the same time. 
It is thereby possible 
that, besides massive primary and secondary components, 
smaller satellites can form  around the primary and/or the secondary.

Among discovered trans-Neptunian binaries, the angular
momentum is usually considerably smaller than the model's
prediction for two identical rarefied preplanetesimals
having the same total mass 
and encountering up to the Hill radius from circular heliocentric orbits. 
The predictions for two preplanetesimals with unequal masses equivalent to
those observed in the trans-Neptunian binaries are also too large. 
Furthermore,
the observed separations between components are usually much smaller than 
the Hill radii. These facts support the hypothesis that most preplanetesimals were already 
partly compressed at the moment of collision, i.e., 
smaller than their Hill radii and/or centrally concentrated. 
The contraction of preplanetesimals could be slower farther from
the Sun, which would explain the greater fraction of binaries
formed at greater distances from the Sun.

The results of this research, which has focussed on the angular momentum of colliding
preplanetesimals, can also be used to analyse the formation
of axial rotation of rarefied and solid bodies.

\section*{ACKNOWLEDGMENTS}
I am thankful to an anonymous reviewer,  Dr. Benjamin Mathiesen, and Bella Nicholls, 
an editorial assistant, for helpful comments.

\label{lastpage}

\end{document}